\documentstyle[aas2pp4]{article}

\received{14 September 1998}
\accepted{20 November 1998}
\slugcomment{Accepted to ApJL, \#985688}

\lefthead{Ishimaru \& Wanajo}
\righthead{ENRICHMENT OF {\it r}-PROCESS ELEMENTS}

\begin{document}

\title{Enrichment of the {\it r}-process Element Europium in the Galactic Halo}

\author{Yuhri Ishimaru}
\affil{Institute of Astronomy, 
       University of Tokyo, 
       2-21-1 Osawa, Mitaka, Tokyo 181-8588, Japan;
       ishimaru@mtk.ioa.s.u-tokyo.ac.jp}

\and

\author{Shinya Wanajo}
\affil{Division of Theoretical Astrophysics, 
       National Astronomical Observatory,
       2-21-1 Osawa, Mitaka, Tokyo 181-8588, Japan;
       wanajo@diamond.mtk.nao.ac.jp}

\begin{abstract}

We investigate the enrichment of europium, as a representative of {\it
r}-process elements, in the Galactic halo. In present chemical evolution
models, stars are assumed to be formed through shock processes by
supernovae (SNe). The enrichment of the interstellar medium is
calculated by a one-zone approach. The observed large dispersions in
[Eu/Fe] for halo stars, converging with increasing metallicity, can be
explained with our models. In addition, the mass range of SNe for the
{\it r}-process site is constrained to be either stars of $8-10 M_\odot$
or $\gtrsim 30 M_\odot$.

\end{abstract}

\keywords{Galaxy: evolution --- Galaxy: halo --- nuclear reactions,
          nucleosynthesis, abundances --- stars: abundances --- stars:
          Population II --- supernovae: general}

\section{Introduction}

Recent abundance analysis of metal-poor halo stars reveals the presence
of large dispersions in heavy elements. This may be interpreted as a
result of incomplete mixing of the interstellar medium (ISM) at the
beginning of the Galaxy (\cite{Gilr88}). Each type of element shows a
unique dispersion, which cannot be simply explained by spatial
inhomogeneity of the ISM. The dispersions of neutron-capture elements
like Sr, Ba, and Eu range $\sim 300$-fold, while those of
$\alpha$-elements and iron-peak elements range typically $\lesssim
10$-fold (\cite{McWi95}; \cite{Ryan96}). This fact implies mixing of
ejecta from small numbers of SNe into the parent clouds (\cite{Audo95};
McWilliam 1997, 1998).

In previous chemical evolution models, observed stellar compositions are
taken to represent those of the ISM when the stars were formed. It may
not be true, however, if star formations are mainly triggered by
SNe. The composition of the formed star must be a mixture of the ISM and
the individual SNe ejecta. Mathews, Bazan, \& Cowan (1992) have examined
the enrichment of neutron-capture elements using a chemical evolution
model of the ISM. They concluded that the delayed increasing of [Eu/Fe]
in terms of [Fe/H] favored origins of low mass SNe ($\sim 7-8 M_\odot$).
However, this might change when including the large dispersion in
[Eu/Fe] for halo stars.

The excellent agreement of the neutron-capture elements in the halo
stars CS 22892-052, HD 115444, HD 122563, and HD 126238 (\cite{Sned96},
1998) with the solar {\it r-}process abundance pattern implies the
presence of one robust {\it r-}process site (\cite{Cowa98}). However,
the origins of {\it r-}process elements are still unknown. Although the
neutrino winds in SNe have been thought to be a promising site, this
scenario involves serious problems, e.g., in obtaining sufficient
entropy (\cite{Taka94}; \cite{Woos94}; \cite{Qian96}). Collapsing
O-Ne-Mg cores are also thought to be an {\it r-}process site
(\cite{Whee98}). The O-Ne-Mg core, as the final stage of an $8-10
M_\odot$ star, may explode by a prompt shock (\cite{Hill84}). The
ejected shell contains a rather low electron fraction region due to
electron captures, which may be a promising {\it r-}process site
(\cite{Meye97}).

In this Letter, we discuss the enrichment of europium in halo stars
formed through stimulated processes by SNe. Details of our models
including other elements will be presented in a forthcoming paper.

\section{Methods}

The evolutions of the ISM in the Galactic halo are calculated by a
one-zone halo model which loses gas through accretion onto the disk. The
star formation and accretion rates are assumed to be proportional to the
gas fraction of the halo.  The star formations obey the Salpeter initial
mass function in the mass range $0.05-60 M_\odot$. The coefficients for
the accretion rate and the star formation rate are adjusted to fit to
the observational data of [O/Fe] versus [Fe/H] (e.g., \cite{Barb88};
\cite{Edva93}) and the metallicity distribution of halo stars
(\cite{Sand87}). Stellar lifetimes are adopted from Schaller et
al. (1992). Type Ia SNe are supposed to occur simply with a lifetime of
2.5 Gyr for $\sim 5\%$ of the $3-8 M_\odot$ stars (\cite{Ishi94};
\cite{Yosh96}).

We assume that star formation is initiated by SNe.  An SN remnant is
supposed to expand spherically until reaching the merge radius with the
ISM (typically $\sim 100$~pc; \cite{Ciof88}). The chemical composition
of a formed star is assumed to be the mass average of the ``snowplowed"
ISM and the SN ejecta. The mass of the SN progenitor is chosen randomly
but obeying the initial mass function. Further, we presume the following
two cases in which the merge radii have a Gaussian distribution in the
logarithmic scale within a factor of 1.5 (case~1) and no distribution
(case~2).

Stellar yields for Type II and Type Ia SNe are taken from Nomoto et
al. (1997a) and Nomoto et al. (1997b), respectively.  The $8-10 M_\odot$
stars are assumed to produce no iron.  For simplicity, we do not take
into account the metallicity effects as examined by Woosley and Weaver
(1995). We do not include yields of the stars smaller than $8 M_\odot$
either, which does not effect the results of the present study. The {\it
r-}process elements are supposed to be produced only in Type II SNe.  In
this study, we examine the following three models (hereafter models 1,
2, and 3) in which europium is produced from the stars: (1) $8-10
M_\odot$ , (2) $\ge 10 M_\odot$, and (3) $\ge 30 M_\odot$.  For each
model, the mass of produced europium is assumed to be constant over the
stellar mass range, scaled to be $[{\rm Eu/Fe}] = 0.5$ in the ISM at
$[{\rm Fe/H}] = -2$.

\section{Enrichment of europium in the halo}

\begin{figure}[t]
\plotone{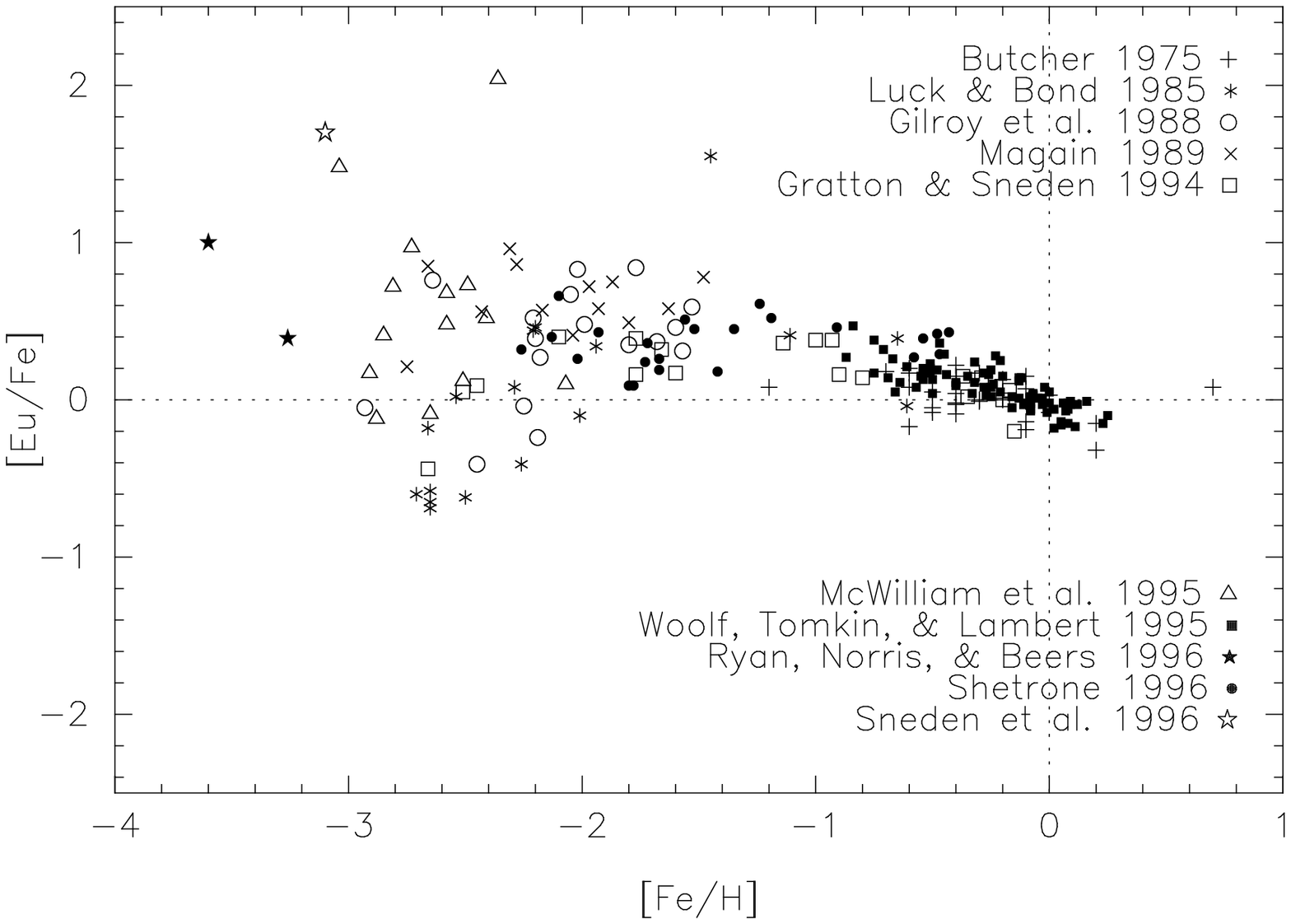}
\caption{\footnotesize
[Eu/Fe] vs. [Fe/H] for halo and disk stars,
from the data of \cite{Butc75}, \cite{Luck85}, Gilroy et al. (1988),
\cite{Maga89}, \cite{Grat94}, McWilliam et al. (1995), \cite{Wool95},
Ryan, Norris, \& Beers (1996), \cite{Shet96}, and Sneden et al. (1996).}
\end{figure}

Figure~1 shows observational data of [Eu/Fe] versus [Fe/H] for halo and
disk stars. Figures~2, 3, and 4 illustrate the enrichment of europium in
the halo by models~1, 2, and 3, respectively. The evolution of the ISM
is represented by a line, and the compositions of stars are plotted by
open stars (case~1) and filled stars (case~2).  As can be seen in the
decrease of [Eu/Fe] for ${\rm [Fe/H]} \gtrsim -1$ in Figures~2--4, the
iron production by Type Ia SNe has little contribution to the enrichment
for metal-poor halo stars.

We see from models~1 and 3 (Figs.~2 and 4) that a large dispersion in
[Eu/Fe] appears for stars at $[{\rm Fe/H}] \sim - 3$, converging towards
the value of the ISM with metallicity. These results are in excellent
agreement with the observations (Fig.~1). The overproduction of europium
in stars compared to the value of the ISM is due to the star formations
by SNe that add europium with little or no iron to the ISM. In contrast,
the underproduction is due to SNe that add iron without europium.  Thus,
for model~1, a significant overabundance in [Eu/Fe] appears for the
stars formed by $8-10 M_\odot$ SNe. For model~3, stars $\ge 30 M_\odot$
are supposed to produce europium with significantly high ratios relative
to iron, since these stars account for only $\sim 15\%$ of all SNe in
number. As a result, a large dispersion also appears.  In contrast to
models~1 and 3, stars in model~2 (Fig.~3) show little dispersion in
[Eu/Fe], since all SNe ($\ge 10M_\odot$) produce iron and europium with
similar ratios. This result is rather close to the observational data of
$\alpha$-elements.

The distribution of merge radii of SN remnants (case~1) can be another
reason for the dispersion in [Eu/Fe]. For case~2, stars show a smaller
scatter due to no distribution in the merge radii, and are thus clearly
separated into overabundant and underabundant ones.

As can be seen in Figures~2 and 4, the number of stars overabundant in
europium is smaller than those that are underabundant. This is also in
good agreement with the observational results (Fig.~1).  The reason is
simply that the number of SNe producing europium is smaller than of
other SNe in these models.

\begin{figure}[t]
\plotone{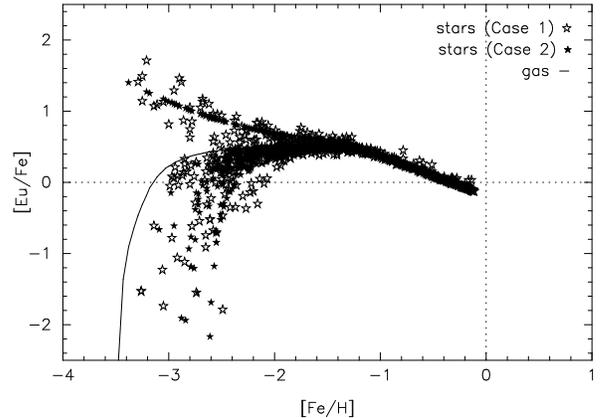}
\caption{\footnotesize
Enrichment of europium as a function of
metallicity in the Galactic halo for model~1. The production site of
europium is assumed to be $8-10 M_\odot$ stars. The evolution of the ISM is
indicated by a line, and the compositions of stars are denoted by
open stars (case~1) and filled stars (case~2).}
\end{figure}

\begin{figure}[t]
\plotone{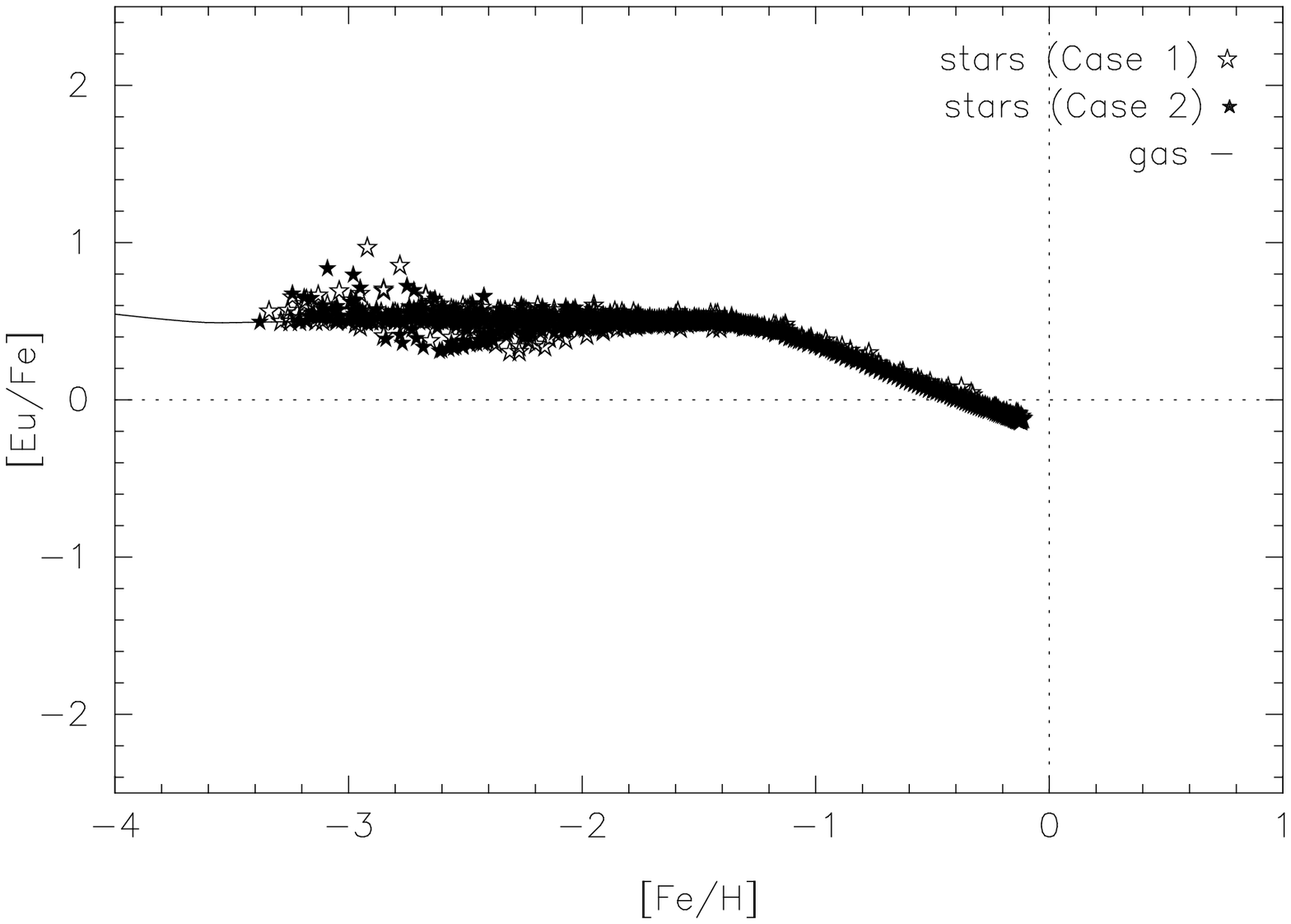}
\caption{\footnotesize
Same as Fig.~2, but the site of
europium is assumed to be $\ge 10 M_\odot$ stars (model~2).}
\end{figure}

\begin{figure}[t]
\plotone{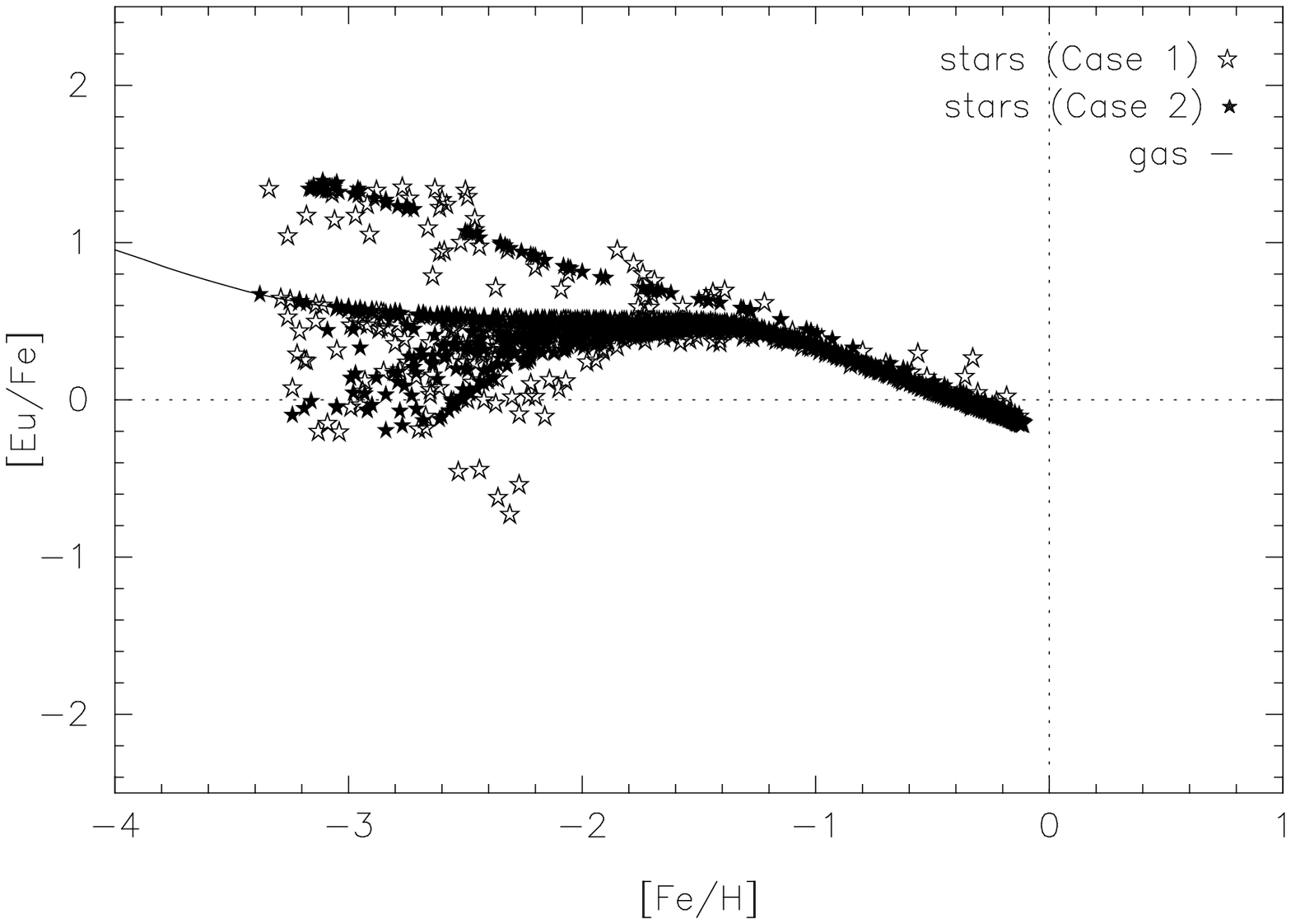}
\caption{\footnotesize
Same as Fig.~2, but the site of
europium is assumed to be $\ge 30 M_\odot$ stars (model~3).}
\end{figure}

\section{Discussion and Conclusions}

In this study we find that a large dispersion in [Eu/Fe] for halo stars
is mainly due to star formation by individual SNe that eject unique
yields. There may be other reasons for the dispersion (e.g., a
distribution of merge radii of SN remnants with the ISM [see \S3] and a
spatial inhomogeneity of the ISM), which are not dealt with in this
study. We should emphasize, however, that the stochastic star formation
process triggered by different masses of SNe is essential to reproduce
the $\sim 300$-fold dispersions in neutron-capture elements.  A
dispersion caused by other factors may be $\sim 10$-fold at most, since
it must appear similarly in other types (e.g., $\alpha$- and iron-peak)
of elements.

The results of our study indicate that the production sites of europium,
as a representative of {\it r-}process elements, must satisfy at least one of
the following two conditions: (1) europium is produced with little iron
or (2) the number of SNe producing europium accounts for only a small
fraction of all SNe. We suggest here a couple of possible sites for
these conditions. The first is the explosion of $8-10 M_\odot$ stars. These
stars are expected to produce little iron (Hillebrandt et al. 1984). In
fact, the progenitor of the Crab Nebula, which shows no significant
enrichment in metal, has been suspected to be an $8-10 M_\odot$ star
(\cite{Nomo82}). The second site is the explosion of stars $\gtrsim 30 M_\odot$.
The subsequent neutrino winds may obtain substantially high entropy
owing to the massive neutron stars (\cite{Qian96}). These stars account
for only $\sim 15\%$ of all SNe. Furthermore, the explosions of these
stars may eject significantly less iron than $\sim 0.1 M_\odot$
adopted in this study (Woosley \& Weaver 1995). In fact, very low masses
of $^{56}$Ni in the ejecta of SN 1994W ($\sim 0.0026 M_\odot$;
\cite{Soll98}) and SN 1997D ($\sim 0.002 M_\odot$; \cite{Tura98}) have
been expected to be due to $25-40 M_\odot$ progenitors.  It should be
noted that coalescing neutron stars may also produce {\it r-}process elements
without iron (\cite{Ross98}).  However, they may not be major sources
for the large enhancement in [Eu/Fe], since their lower kinetic energies
are not enough to trigger star formations.

In the light of the observational results (Fig.~1), model~1 seems more
likely for the following two reasons. First, in model~1 some stars show as
high an [Eu/Fe] as $\sim 1.7$ at $[{\rm Fe/H}] \sim - 3$. These stars can
be a possible explanation for CS 22892-052. However, model~3 would
produce stars with a significantly larger dispersion in [Eu/Fe] if
smaller iron yields were adopted for stars $\geq 30 M_\odot$. Second, in
model~1 there is no sharp peak in the distribution of stars for $[{\rm
Fe/H}] \lesssim -2$ (case~1) that is in good agreement with the
observations. In model~3 the concentration of stars on the line of the
ISM for $[{\rm Fe/H}] \lesssim -2$ is due to the star formations by
$8-10 M_\odot$ stars which add neither iron nor europium. It should be
noted, however, that the question whether $8-10 M_\odot$ stars explode
or not is still open. Another outstanding difference between these
models is that model~1 has stars in [Eu/Fe] down to $\sim -2$ that have
not detected by observations yet. This is a consequence of the delayed
increasing of europium in the ISM due to lower mass progenitors. Hence,
future surveys of metal-poor halo stars for low [Eu/Fe] will be
important to distinguish the above two sites for the {\it r-}process. An
alternative way to distinguish these sites may be to apply our models to
other neutron-capture elements like Sr and Ba which have much more data
than Eu. In addition, the decreasing trend of [Ba/Fe] towards lower
metallicity may support $8-10 M_\odot$ progenitors. The future direction
of this study will be one that investigates enrichment of a number of
neutron-capture elements consistently.

\acknowledgments

We would like to acknowledge useful discussions with 
T. Young, S. Ryan, and T. Kajino.
This work was supported in part by Japan
Society for Promotion of Science.


\clearpage

\begin{thebibliography}{}

\bibitem[Audouze \& Silk 1995]{Audo95}
 Audouze, J. \& Silk, J. 1995, \apjl, 451, 49
\bibitem[Barbuy 1988]{Barb88}
 Barbuy, B. 1988, \aap, 191, 121
\bibitem[Butcher (1975)]{Butc75}
 Butcher, H. R. 1975, \apj, 199, 710
\bibitem[Cioffi, McKee, \& Bertschinger 1988]{Ciof88}
 Cioffi, D. F., McKee, C. F., \& Bertschinger, E. 1988, \apj, 334, 252
\bibitem[Cowan et al. 1998]{Cowa98}
 Cowan, J. J., Pfeiffer, B., Thielemann, F. -K., Sneden, C., 
 Burles, S., Tytler, D., \& Beers, T. C. 1998, \apj, submitted 
 and astro-ph/9808272
\bibitem[Edvardsson et al. 1993]{Edva93}
 Edvardsson, B., Andersen, J., Gustafsson, B., Lambert, D. L., Nissen, P. E.,
 \& Tomkin, J. 1993, \aap, 275, 101
\bibitem[Gilroy et al. 1988]{Gilr88}
 Gilroy, K. K., Sneden, C., Pilachowski, C. A., \& Cowan, J. J. 1988, 
 \apj, 327, 298
\bibitem[Graton \& Sneden (1994)]{Grat94}
 Graton, R. G. \& Sneden, C. 1994, \aap, 287, 927
\bibitem[Hillebrandt, Nomoto, \& Wolff 1984]{Hill84}
 Hillebrandt, W., Nomoto, K., \& Wolff, R. G. 1984, \aap, 133, 175
\bibitem[Ishimaru 1994]{Ishi94}
 Ishimaru, Y. 1994, Master's thesis, Univ. of Tokyo
\bibitem[Luck \& Bond (1985)]{Luck85}
 Luck, R. E. \& Bond, H. E. 1985, \apj, 292, 559
\bibitem[Magain (1989)]{Maga89}
  Magain, P. 1989, \aap, 209, 211
\bibitem[Mathews, Bazan, \& Cowan 1992]{Math92}
 Mathews, G. J., Bazan, G., \& Cowan, J. J. 1992, \apj, 391, 719
\bibitem[Meyer \& Brown 1997]{Meye97}
 Meyer, B. S. \& Brown, J. 1997, \apjs, 112, 199
\bibitem[McWilliam et al. 1995]{McWi95}
 McWilliam, A., Preston, G. W., Sneden, C., \& Searle, L. 1995, \aj, 109, 2757
\bibitem[McWilliam 1997]{McWi97}
 McWilliam, A. 1997, \araa, 35, 503
\bibitem[McWilliam 1998]{McWi98}
 McWilliam, A. 1998, \aj, 115, 1640
\bibitem[Nomoto et al. 1982]{Nomo82}
 Nomoto, K., Sparks, W. M., Fesen, R. A., Gull, T. R., Miyaji, S.,
 \& Sugimoto, D. 1982, \nat, 299, 803
\bibitem[Nomoto et al. 1997a]{Nomo97a}
 Nomoto, K., Hashimoto, M., Tsujimoto, T., Thielemann, F. -K., 
 Kishimoto, N., Kubo, Y., \& Nakasato, N. 1997a, \nphysa, 616, 79
\bibitem[Nomoto et al. 1997b]{Nomo97b}
 Nomoto, K., et al. 1997b, \nphysa, 621, 467
\bibitem[Qian \& Woosley 1996]{Qian96}
 Qian, Y. -Z. \& Woosley, S. E. 1996, \apj, 471, 331
\bibitem[Ryan, Norris, \& Beers 1996]{Ryan96}
 Ryan, S. G., Norris, J. E., \& Beers, T. C. 1996, \apj, 471, 254
\bibitem[Rosswog et al. 1998]{Ross98}
 Rosswog, S., Liebendoerfer, M., Thielemann, F. -K., Davies, M. B.,
 Benz, W., \& Piran, T. 1998, \aap, in press (astro-ph/9804332)
\bibitem[Sandage \& Fouts 1987]{Sand87}
 Sandage, A. \& Fouts, G. 1987, \aj, 97, 74
\bibitem[Schaller et al. 1992]{Scha92}
 Schaller, G., Schaerer, D., Meynet, G., \& Maeder, A. 1992, A\&AS, 96, 269
\bibitem[Shetrone (1996)]{Shet96}
 Shetrone, M. D. 1996, \aj, 112, 1517
\bibitem[Sneden et al. 1996]{Sned96}
 Sneden, C., McWilliam, A., Preston, G. W., Cowan, J. J., Burris, D. L., 
 \& Armosky, B. J. 1996, \apj, 467, 819
\bibitem[Sneden et al. 1998]{Sned98}
 Sneden, C., Cowan, J. J., Debra, L. B., \& Truran, J. W. 1998, \apj, 496, 235
\bibitem[Sollerman, Cumming, \& Lundqvist 1998]{Soll98}
 Sollerman, J., Cumming, R. J., \& Lundqvist, P. 1998, \apj, 493, 933
\bibitem[Takahashi, Witti, \& Janka 1994]{Taka94}
 Takahashi, K., Witti, J., \& Janka, H. -Th. 1994, \aap, 286, 857
\bibitem[Turatto et al. 1998]{Tura98}
 Turatto, M., and et al. 1998, \apjl, 498, 129
\bibitem[Wheeler, Cowan, \& Hillebrandt 1998]{Whee98}
 Wheeler, J. C., Cowan, J. J., \& Hillebrandt, W. 1998, \apjl, 493, 101
\bibitem[Woolf, Tomkin, \& Lambert (1995)]{Wool95}
 Woolf, V. M., Tomkin, J. \& Lambert, D. L. 1995, \apj, 453, 660
\bibitem[Woosley et al. 1994]{Woos94}
 Woosley, S. E., Wilson, J. R., Mathews, G. J., Hoffman, R. D., \&
 Meyer, B. S. 1994, \apj, 433, 229
\bibitem[Woosley \& Weaver 1995]{Woos95}
 Woosley, S. E. \& Weaver, T. A. 1995, \apjs, 101, 181
\bibitem[Yoshii, Tsujimoto, \& Nomoto 1996]{Yosh96}
 Yoshii, Y., Tsujimoto, T., \& Nomoto, K. 1996, \apj, 462, 266

\end{thebibliography}
\end{document}